\documentclass[12pt,fleqn]{article}
\usepackage[DIV12]{typearea}
\usepackage{amsmath,amsfonts,amssymb}
\usepackage{graphicx}
\usepackage{cite}
\usepackage{mathrsfs}
\usepackage{bbm}
\usepackage{units}
\usepackage[small,loose]{subfigure}  

\newcommand{\eVdist}{\kern-0.06em}

\newcommand{\gev}{\:\text{Ge\eVdist V}}
\newcommand{\tev}{\:\text{Te\eVdist V}}

\DeclareMathOperator{\sign}{sign}

\newcommand{\SU}[1]{\ensuremath{\mathrm{SU}(#1)}}

\allowdisplaybreaks[1]

\title{Precision gauge unification in the MSSM}

\begin{document}

\begin{titlepage}

\begin{flushright}
OHSTPY-HEP-T-09-004\\
DOE/ER/01545-883\\
TUM-HEP 739/09
\end{flushright}

\vspace*{1.0cm}

\begin{center}
{\Large\bf
Precision gauge unification in the MSSM
}

\vspace{1cm}

\textbf{
Stuart Raby\footnote[1]{Email: \texttt{raby@pacific.mps.ohio-state.edu}}{}$^a$,
Michael Ratz\footnote[2]{Email: \texttt{mratz@ph.tum.de}}{}$^b$,
Kai Schmidt-Hoberg\footnote[3]{Email: \texttt{kai.schmidt-hoberg@ph.tum.de}}{}$^b$
}
\\[5mm]
\textit{\small
{}$^a$ Department of Physics, The Ohio State University,\\
191 W. Woodruff Ave, Columbus, OH 43210, USA
}
\\[3mm]
\textit{\small
{}$^b$ Physik-Department T30, Technische Universit\"at M\"unchen, \\
~~James-Franck-Stra\ss e, 85748 Garching, Germany
}
\end{center}

\vspace{1cm}

\begin{abstract}
We discuss the issue of precision gauge unification in the MSSM. We
find that a comparably light gluino, as it emerges in certain
patterns of soft supersymmetry breaking, can be a key ingredient for
ensuring precision gauge unification without relying on the presence
of extra particles around the scale of grand unification. In
particular, the so-called mirage pattern for gaugino masses can
naturally lead to precision gauge unification. There is also an
interesting correlation with reduced fine-tuning, due to rather
light gluinos.
\end{abstract}

\end{titlepage}

\newpage

\section{Introduction}

It is well known that gauge coupling unification
\cite{Dimopoulos:1981yj,Dimopoulos:1981zb,Ibanez:1981yh,Einhorn:1981sx,Marciano:1981un} looks
very promising in the minimal supersymmetric extension of the standard model,
the MSSM. A precise analysis reveals that, under the assumption of a
`standard' supersymmetry breaking scenario, gauge couplings do not meet
precisely but in many cases the strong fine structure constant
$\alpha_3=g_3^2/(4\pi)$ turns out to be about $3\,\%$ smaller than $\alpha_1$
and $\alpha_2$ at  $M_\mathrm{GUT}$, which is defined as the scale where
$\alpha_1$ and $\alpha_2$ unify. This happens in scenarios in which
\begin{itemize}
 \item gaugino masses unify at $M_\mathrm{GUT}$;
 \item scalar masses are universal.
\end{itemize}
One can correct for the discrepancy in many ways, most plausably
through threshold corrections at the TeV or at the GUT scale (see,
e.g., \cite{Alciati:2005ur}; see also \cite[S.~Raby,`Grand unified
theories']{Amsler:2008zzb}). Such thresholds may stem from exotics
around the GUT scale (possibly from the GUT breaking
sector) or, as in orbifold GUTs, 
from Kaluza-Klein modes between the compactification and cut-off scales
(see e.g.\ \cite{Ibanez:1991zv,Ibanez:1992hc,Nilles:1995kb,Hall:2001pg}).
In either case, such corrections will always exist.
 However, depending on the mechanism of GUT breaking they can be
highly suppressed (see e.g.\ \cite{Hebecker:2004ce} for a discussion).
 In this letter, we will focus on the question of how to obtain precision gauge
coupling unification in the MSSM, i.e.\ with a subdominant
contribution from extra states.

One may attribute the `non-unification' to the fact that $\alpha_3$
runs too fast in scenarios that have the above two properties. In
order to slow down the renormalization group (RG) evolution of
$\alpha_3$ one might therefore lower the masses of colored states.
Perhaps the simplest option for such a colored particle is the
gluino, which transforms as an $\boldsymbol{8}$-plet under
$\SU3_\mathrm{C}$; lowering the masses of triplet-anti-triplet pairs
often leads to problems with proton decay. 
Specifically, in scenarios in which the gaugino masses are
non-universal at (or slightly below) the GUT scale, the deviation
$\epsilon_3$ from the other two $\alpha$s can vanish. 

Let us note that, in general, the threshold effects at low energies seem to
be more important than those at the high scale. That is, any correction to
the fine-structure constants will be of the form
\begin{equation}
    \Delta \alpha_i~=~b_i'\, \alpha_i^2 \ln(m/M_*)\;,
\end{equation}
where $b_i'$ is a `color factor', $m$ is the mass of the particle, $M_*$ is either
$M_\mathrm{GUT}$ or $M_\mathrm{SUSY}$, and $\alpha_i$ is the fine structure
constant at $M_*$. Since $\alpha_3$ is much bigger at low energies, in order to
get the same numerical correction one has to have a much larger ratio $m/M_*$ at
high energies than at low energies. We will hence focus on settings in which the
gluino is lighter than in the above-mentioned `standard' scenarios. 

It has been known for some time that a comparatively light gluino allows us to
alleviate the tension between the predicted and measured values of $\alpha_3$ 
\cite{Carena:1993ag,Roszkowski:1995cn};
the main point of our analysis is that there are in fact well-motivated patterns of supersymmetry breaking
which lead to precision gauge unification.
It is worthwhile to stress that non-universal boundary conditions
for the gaugino masses can be consistent with unification; they
arise, for instance, in the scheme of mirage mediation
\cite{LoaizaBrito:2005fa}, which has been first discussed in the
context of type II B flux compactifications
\cite{Choi:2005ge,Choi:2005uz,Falkowski:2005ck}, but may also occur
in heterotic string theory \cite{Lowen:2008fm}. In particular,
non-universal gaugino masses are something that one expects in
settings in which dominant supersymmetry breaking comes from a field
that does not enter the gauge kinetic function, e.g.\ a matter field
\cite{Lebedev:2006qq,Lebedev:2006qc}. There, the interplay between a
suppressed tree-level term and quantum corrections render the
gaugino masses non-universal slightly below the GUT scale. These
corrections turn out to lower the gluino mass with respect to the
other gaugino masses, which is precisely what is needed to achieve
precision gauge coupling unification without the need of invoking
high-energy thresholds.

This letter is organized as follows: in the next section we briefly discuss the
low energy thresholds which are needed for precision gauge coupling unification
before we come to specific scenarios in section~\ref{sec:scenarios}.
In section~\ref{sec:Pheno} we comment on phenomenological properties of settings
with precision unification.
Section~\ref{sec:summary} contains our summary.

\section{Low energy thresholds}

Before we turn to specific scenarios with motivated high-energy boundary conditions for the soft terms,
let us briefly consider the main low energy thresholds which are needed in order to obtain
precision gauge coupling unification.
The picture here is that at low energies the couplings evolve according to the
renormalization group equations of the standard model (SM) until the superpartners kick in at
about a TeV and modify the running.
In the qualitative discussion of this section,
we will restrict ourselves to the 1-loop level,
while for the numerical analyses in the next section we utilize standard 2-loop codes.
As eluded to already in the introduction, one may attribute the `non-unification' to the fact that
$\alpha_3$ runs too fast in the SM, resulting in
\begin{equation}\label{eq:Deviation}
 \epsilon_3~:=~\frac{\alpha_3-\alpha_{1,2}}{\alpha_{1,2}} ~\simeq~ -0.03
\end{equation}
at $M_\mathrm{GUT}$, which is defined to be the scale at which $\alpha_1$ and
$\alpha_2$ unify.
If we insisted on unification and enforced $\epsilon_3=0$ in such a setting, 
$\alpha_3$ would be off from its measured value at the weak scale by roughly 
$\tfrac{\Delta \alpha_3(M_Z)}{\alpha_3 (M_Z)} 
\sim \tfrac{\alpha_3(M_Z)}{\alpha_3(M_\text{GUT})} \epsilon_3$
so that a 3\% error on $\epsilon_3$ would correspond to $\tfrac{\Delta \alpha_3(M_Z)}{\alpha_3 (M_Z)} \sim 10\%.$

To see the impact of the low energy thresholds on $\epsilon_3$ let us have a closer look at the one-loop
renormalization group equations (RGEs) for the gauge couplings $g_i$,
\begin{equation}
\frac{\text{d}g_i}{\text{d}t} ~\equiv~ \beta(g_i)
~=~ \frac{b_i}{16\pi^2} T_i\,  g_i^3\;, \quad (b_1,b_2,b_3)=(\tfrac{33}{5},1,-3) \;,
\end{equation}
where $t\equiv \text{ln}(q/q_0)$ with $q$ the renormalization scale. Here the $T_i$ take into account the
thresholds from the standard model superpartners. They are given by~\cite{Lahanas:1994dj}
\begin{subequations}\label{eq:theta}
\begin{eqnarray}
 T_1
 & = &
 \frac{1}{33}\left(20 + \theta_{\widetilde{H}_1} + \theta_{\widetilde{H}_2}
  +\frac{1}{2}\left( \theta_{H_1} + \theta_{H_2}\right)
  \right.\nonumber\\
 & &  \hphantom{\frac{1}{33}\left(\right.}\left.{}
 +\sum_{i=1}^3\left( \tfrac{1}{2}  \theta_{\widetilde{L}_i}
 + \theta_{\widetilde{E}_i}  + \tfrac{1}{6}  \theta_{\widetilde{Q}_i}  + \tfrac{4}{3}  \theta_{\widetilde{U}_i}
 + \tfrac{1}{3}  \theta_{\widetilde{D}_i}   \right)    \right)\;, \\
 T_2
 & = &
 -\frac{10}{3}+\frac{4}{3}\theta_{\widetilde{W}}+\frac{1}{3}\left(\theta_{\widetilde{H}_1}
 +\theta_{\widetilde{H}_2}\right)  +\frac{1}{6}\left( \theta_{H_1} + \theta_{H_2}\right)
 + \frac{1}{6}\sum_{i=1}^3\left(3 \theta_{\widetilde{Q}_i} +
 \theta_{\widetilde{L}_i}\right)\;,\\
 T_3
 & = & \frac{7}{3}-\frac{2}{3} \theta_{\widetilde{G}} -\frac{1}{18}\sum_{i=1}^3\left(2 \theta_{\widetilde{Q}_i}
 + \theta_{\widetilde{D}_i}+ \theta_{\widetilde{U}_i}\right) \;,
\end{eqnarray}
\end{subequations}
with the theta function $\theta_\phi=\theta(q^2-m_\phi^2)$ accounting for the new contribution of the particle $\phi$
with mass $m_\phi$. Above all thresholds the $T_i$ become unity such that we encounter the usual MSSM running,
while below all thresholds the $T_i$ take values such that $(b_1 T_1,b_2 T_2,b_3 T_3)=(4,-10/3,-7)$.
When we include one Higgs we obtain $(41/10,-19/6,-7)$ as in the standard model.

Concentrating on $g_3$ for now, we see that to slow down the running it is
easiest to have a light gluino such that the corresponding threshold is crossed
at rather low energies.
As can be seen in \eqref{eq:theta}, also the other colored particles can potentially slow down the running of $\alpha_3$.
However, they also contribute to the running of $\alpha_{1,2}$. To be precise, at one loop, the threshold contributions
$T_i^\text{squarks} b_i$ of the squarks to the three gauge couplings are $(11/10,3/2,2)$.

Let us assume that scalar masses are universal and heavy at the GUT scale. If
the gaugino masses are somewhat lighter, the scalar masses will also be nearly
universal at the electroweak scale and will not induce differential running
between the gauge couplings.\footnote{We neglect here the small contributions
from the Higgs fields.} Then the differential running is governed by the
gauginos only. Hence whether or not we have precision gauge coupling unification
crucially depends on the mass ratio of the gauginos. In particular, as the bino
does not contribute to the running, only the masses of the wino and gluino will
be relevant.

Both, a light gluino and a heavy wino can lead to precision gauge coupling unification by changing the
running of $\alpha_3$ and/or $\alpha_2$ respectively.
In the following we will study scenarios of supersymmetry breaking which result in a
mass pattern allowing for precision gauge coupling unification.

\section{Scenarios}
\label{sec:scenarios}

As we have seen in the previous section the ratios of gaugino masses
are essential ingredients for achieving precision gauge
coupling unification. Therefore let us now consider models which
give non-universal gaugino masses. Non-universal gaugino masses may
arise if the gauge kinetic function $f_{\alpha \beta}$ has some
non-trivial gauge structure. There are several known mechanisms for
obtaining such structures even in grand unified theories or in
string theory.  In the following we shall discuss three examples.

\subsection{Mirage mediation and matter domination}

\begin{figure}[h!]
\centerline{\subfigure[]{\includegraphics[width=7.5cm]{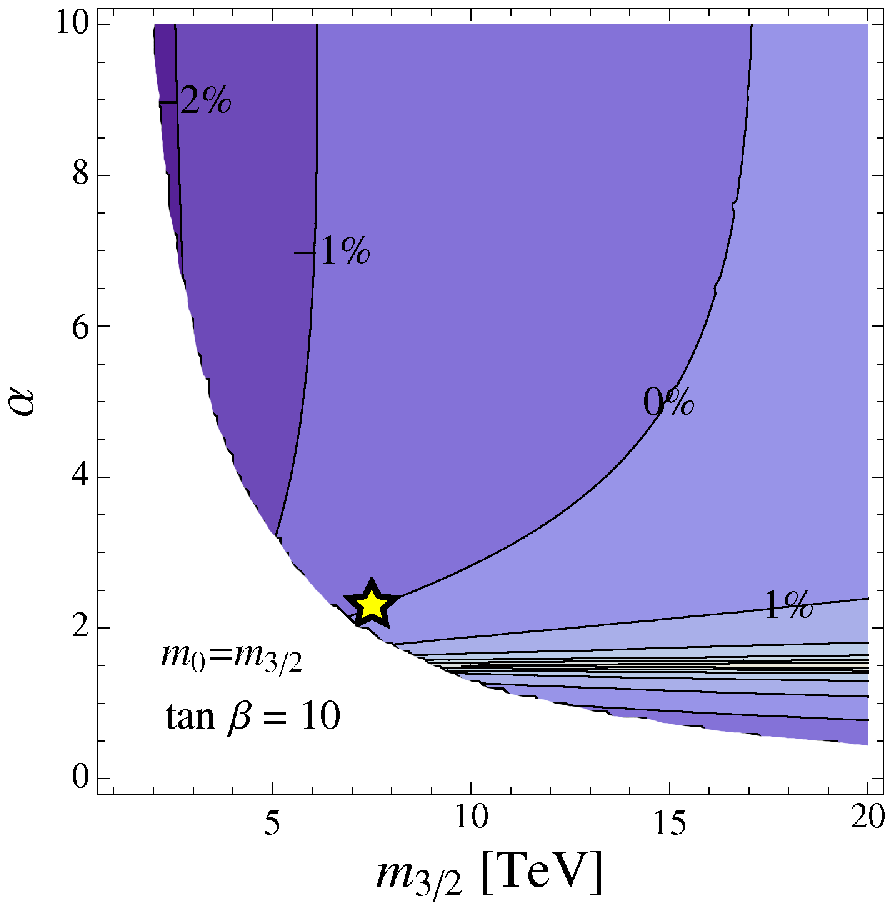}}
\subfigure[]{\includegraphics[width=7.6cm]{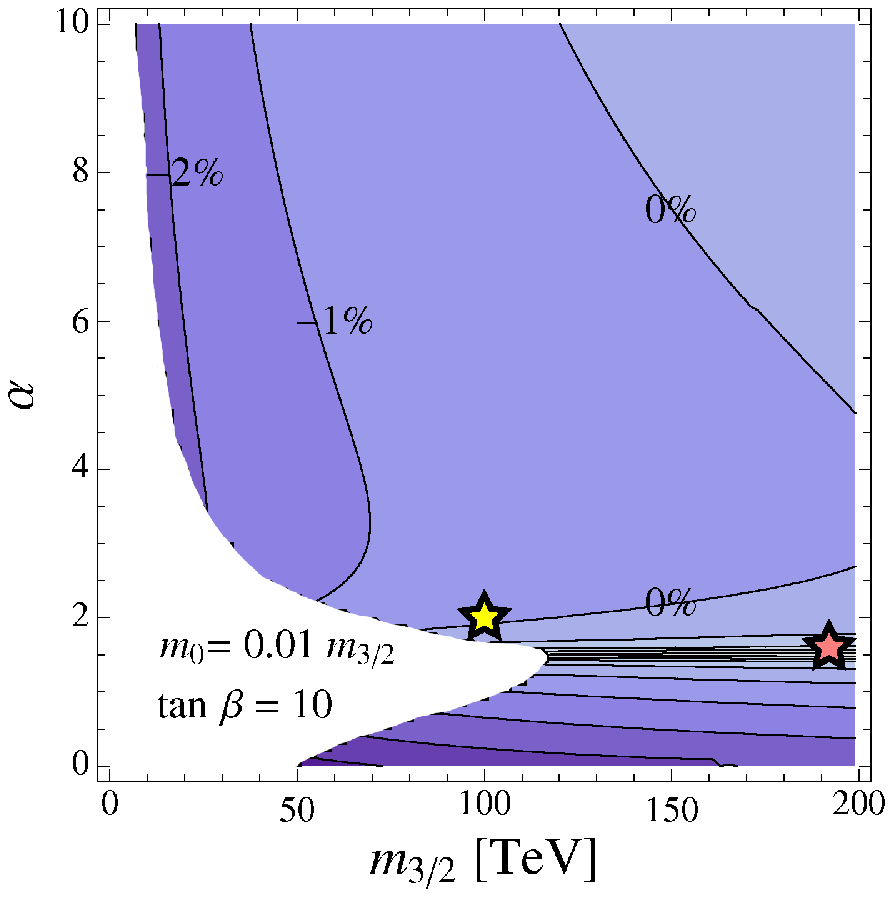}}}
\centerline{\subfigure[]{\includegraphics[width=7.5cm]{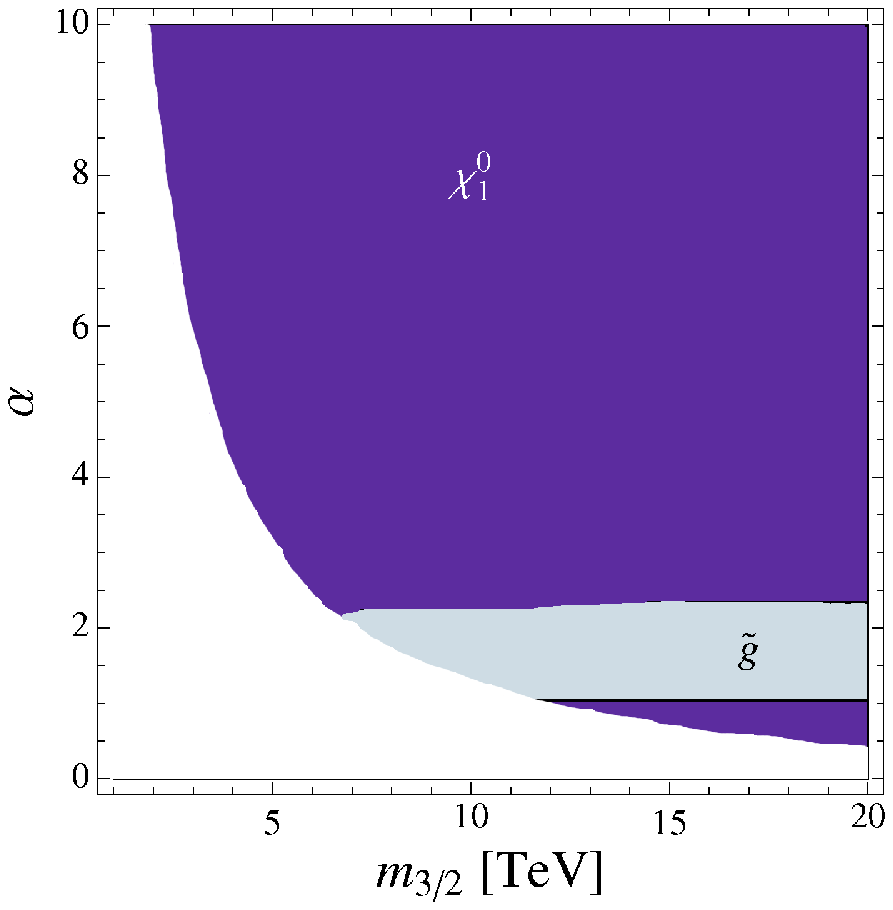}}
\subfigure[]{\includegraphics[width=7.6cm]{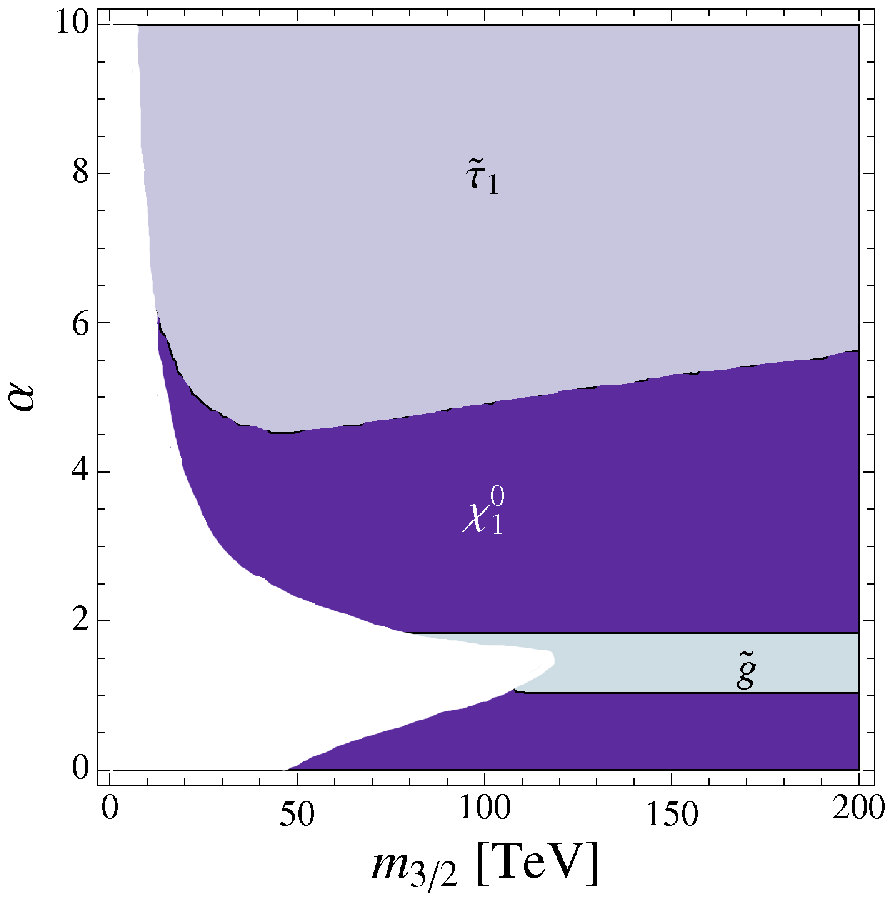}}}
\caption{In panel (a) and (b) we show contours of different $\epsilon_3$ in the
         case of 'matter domination' for $\tan\beta=10$ and $m_0=m_{3/2}$ (a)
         and $m_0=0.01\, m_{3/2}$ (b). The white region is excluded.                                                               The yellow and pink stars denote points in parameter space for which we show the spectra in 
         table~\ref{tab:SampleSpectrum} and table~\ref{tab:SampleSpectrume2}. In panel (c)
         and (d) the corresponding LSPs are shown with $\chi_1^0$ the lightest
         neutralino, $\widetilde{g}$ the gluino and $\widetilde{\tau}_1$ the lighter stau.
}
\label{fig:matter}
\end{figure}

One scenario in which precision gauge coupling unification can be realized is `mirage mediation'.
Here the moduli entering the gauge kinetic function
have suppressed $F$-terms such that the direct contribution to the gaugino
masses and quantum corrections are of the same order.
The boundary conditions for the gaugino masses at the GUT scale read~\cite{Falkowski:2005ck}
\begin{equation}
\label{eq:MiragePatternGaugino}
 M_i ~=~ \frac{m_{3/2}}{16\,\pi^2}\,\left(\alpha+b_i\,g_i^2\right)\;,
\end{equation}
where the $b_i$ denote the one-loop $\beta$-function coefficients and $\alpha$
is a continuous positive parameter.
While the mirage mass relations for gauginos are rather robust, the pattern of masses for the
scalars is more model dependent (cf.~\cite{Choi:2007ka}). In fact, in situations
in which supersymmetry is broken by a matter field, the so-called `matter
domination' scheme, the mirage pattern of the gaugino masses
\eqref{eq:MiragePatternGaugino} is preserved (up to K\"ahler
corrections~\cite{Lowen:2009nr}) while the scalars attain large
positive masses.

Our ansatz for the soft
parameters at the GUT scale is (cf.\ the general formulae in
\cite{Lebedev:2006qq,Lebedev:2006qc})
\begin{subequations}
\begin{eqnarray}
 M_i & = & \frac{m_{3/2}}{16\,\pi^2}\,\left(\alpha+b_i\,g_i^2\right)\;,\\
 m_0^{\boldsymbol{16}} &= & m_{3/2}\;,\\
 A & = & m_0^\mathrm{Higgs}~=~0\;.
\end{eqnarray}
\end{subequations}
The continuous parameters in this scheme are
\[m_{3/2}\;,~\alpha~\text{and}~\tan\beta\;;\]
we fix $\sign\mu$ to be $+1$.
We conducted a scan over the continuous parameters
$m_{3/2},~\alpha$ and $\tan\beta$.
It turns out that the deviation $\epsilon_3$ is almost independent of
$\tan\beta$. In figure~\ref{fig:matter} we therefore show $\epsilon_3$ for
varying $m_{3/2}$ and $\alpha$ with fixed $\tan\beta=10$. The white regions are
excluded as follows: we take into account all points which are invalid according
to SOFTSUSY \cite{Allanach:2001kg}  as well as collider bounds on sparticle
masses.\footnote{Here we require that the  lightest Higgs mass
be above  $112 \gev$ rather than above $114.4 \gev$, since the theoretical error
on the Higgs mass is about $2 - 3 \gev$. Further, we are using 
version 3.0.9 of SOFTSUSY with a top quark mass of
$173.1\gev$ and $\alpha_s=0.1176$.}
This implies a minimal value of $m_{3/2}\gtrsim 7 \tev$
in order to be consistent with precision gauge coupling unification.
At smaller values of $m_{3/2}$ the lighter chargino is below its experimental limit.
Also shown in figure~\ref{fig:matter} are the corresponding LSPs.

To see the impact of smaller $m_0$ let us also consider the case where not only
the gaugino masses but also the scalar masses are suppressed against the
gravitino mass, e.g.\ $m_0^{\boldsymbol{16}}=\tfrac{1}{100} m_{3/2}$. The
corresponding plots are also shown in  figure~\ref{fig:matter}. Sample spectra
for the two cases can be found in table~\ref{tab:SampleSpectrum}.

While this paper is mainly concerned with precision gauge coupling unification, 
there are circumstances under which positive values of $\epsilon_3$ might also
be interesting. Let us briefly discuss an example. If one is to accommodate
negative $\epsilon_3$ in a concrete model, as it occurs in the `usual'
scenarios, one needs to have colored states lighter than the GUT scale. The most
straightforward possibility for such states are the Higgs triplet partners (see
e.g.\ \cite{Dundee:2008ts}). However, there are tight constraints on the masses
of such states coming from proton decay as such triplets induce dangerous
dimension five operators,
which disfavor this
possibility.  On the other hand, a positive $\epsilon_3$ would allow for color
triplets which are substantially heavier than $M_\mathrm{GUT}$, 
thus ameliorating the problems with dimension five operators. 
For example in SU(5) the contribution to $\epsilon_3$ of the color triplet Higgs is given by
\cite{Lucas:1995ic}
\begin{equation}
{\epsilon}_3^\text{Higgs}= {3 {\alpha_\mathrm{GUT}} \over 5\pi}\log \Big|{
 \widetilde{M_t} \over M_\text{GUT}} \Big| 
\end{equation}
with $\widetilde{M_t}$ the effective mass of the Higgs triplet.
Hence for positive $\epsilon_3$ the triplet Higgs can naturally be heavier than $M_\text{GUT}$
(see e.g.\ \cite{Dermisek:2000hr} for a discussion on proton 
stability and an upper bound on  $\widetilde{M_t}$).
In scenarios with positive $\epsilon_3$ the
gluino becomes even lighter. Sample spectra with $\epsilon_3=0.02$ are shown in
table~\ref{tab:SampleSpectrume2}.

\subsection{Non-singlet $\boldsymbol{F}$-terms}

\begin{figure}[h!]
\centerline{\subfigure[]{\includegraphics[width=7.5cm]{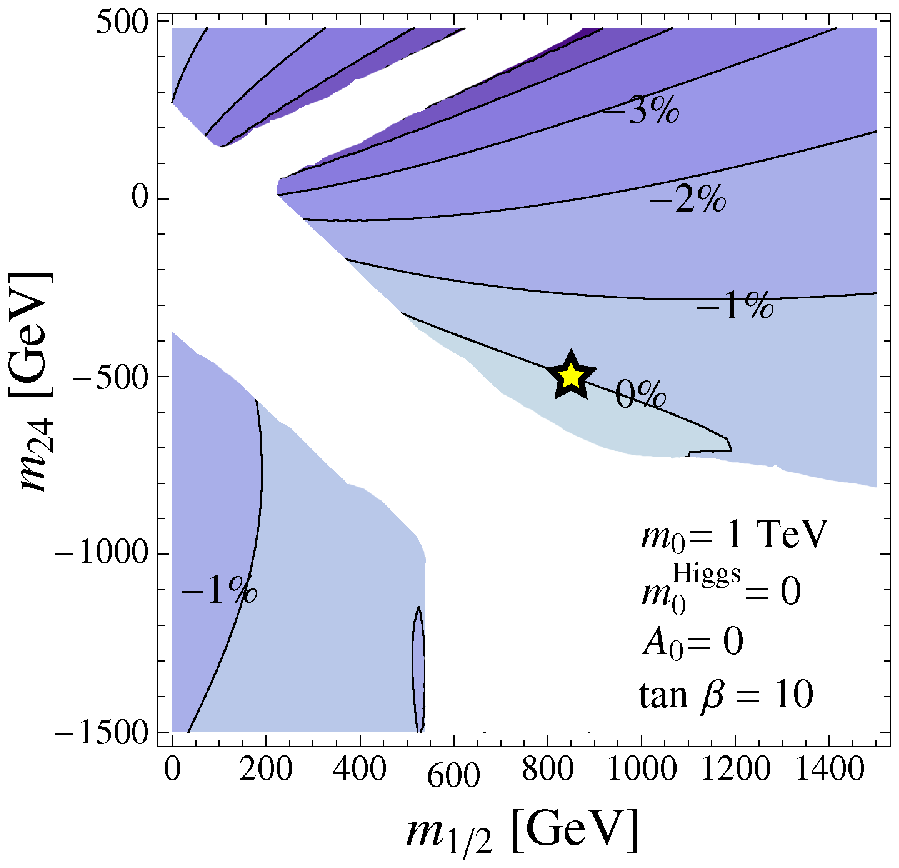}}
\subfigure[]{\includegraphics[width=7.5cm]{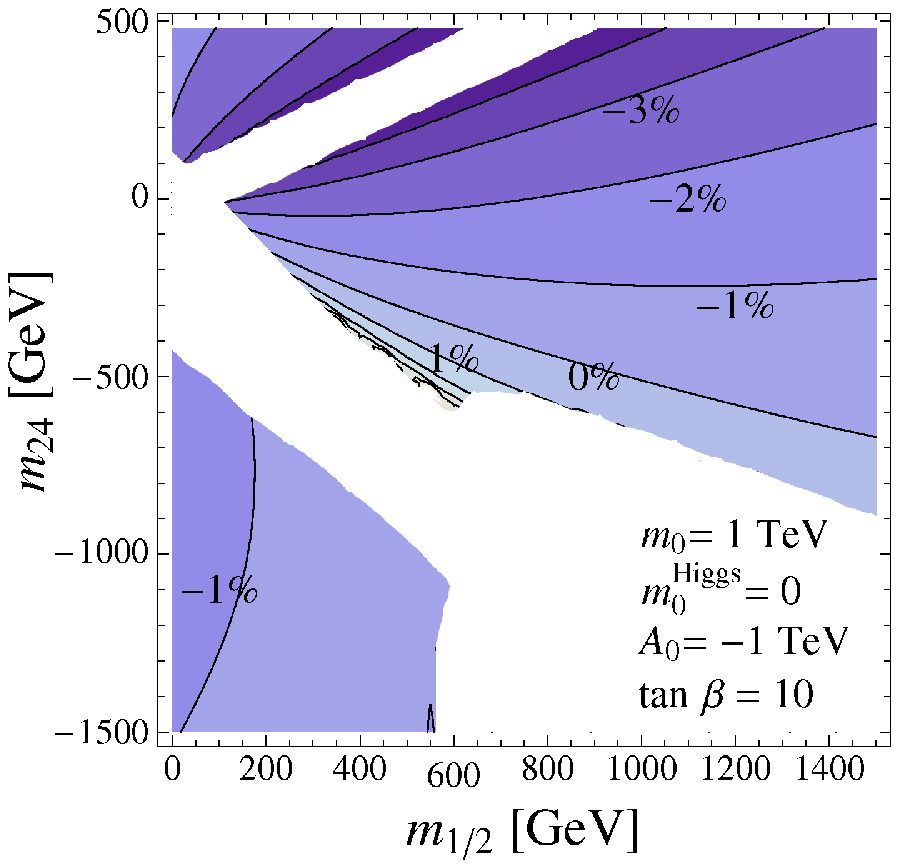}}}
\caption{Here we show contours of different $\epsilon_3$ in the
case of non-singlet $F$-terms.
The white region is excluded. In the allowed regions the lightest neutralino is the LSP.
The yellow star denotes a point in parameter space for which we show the spectrum in table~\ref{tab:SampleSpectrum}.
}
\label{fig:ContoursNUGM}
\end{figure}

Another possible source for non-universal gaugino masses is to have non-singlet chiral fields
which break supersymmetry by $F$-term vacuum expectation values.
These fields have to  transform as the symmetric product of two adjoint
representations of the GUT group, but not necessarily as  singlets.
For concreteness we focus here on SU(5) GUTs. In the
following we concentrate on the two smallest possible representations for the
supersymmetry  breaking fields, which are simply the singlet and the
$\boldsymbol{24}$-plet.
Including also the $\boldsymbol{75}$ and $\boldsymbol{200}$ would of course widen the available
parameter space. Similar relations hold for SO(10).

The high-scale mass patterns of the gauginos of SU(3)$_C$, SU(2)$_\mathrm{L}$
and U(1)$_Y$ turn out to be given as linear combinations of singlet ($m_{1/2}$)
and $\boldsymbol{24}$-plet ($m_{24}$) contributions \cite{Huitu:1999vx},
\begin{eqnarray}
\label{eq:NUGMrelation}
 M_1&=&m_{1/2}-\tfrac{1}{2}\,m_{24} \;,\nonumber\\
 M_2&=&m_{1/2}-\tfrac{3}{2}\,m_{24} \;,\nonumber\\
 M_3&=&m_{1/2}+m_{24}      \;.\label{eq:GUTGauginoMasses124}
\end{eqnarray}

In figure~\ref{fig:ContoursNUGM} we show $\epsilon_3$ as a function of the two gaugino mass
parameters $m_{1/2}$ and $m_{24}$ for fixed $m_0$, $m_0^\text{Higgs}$, $A_0$ and $\tan\beta$.
We find that in the allowed regions $-0.04 \lesssim \epsilon_3 \lesssim 0.03$. 
The white regions are again excluded. The excluded regions are mainly due to the Higgs LEP bound.
A sample spectrum is shown in table~\ref{tab:SampleSpectrum}.

\subsection{GMSB with Higgs messenger mixing}

As a last example of a supersymmetry breaking mechanism which can lead to precision gauge coupling unification
let us look at a particular version of gauge mediation \cite{Raby:1997pb,Raby:1997bpa,Raby:1998xr}.
The boundary conditions at the messenger scale $M\sim M_\text{GUT}$ are determined by five SUSY breaking parameters,
$\Lambda\sim 10^5\gev$, $a\sim b \sim 0.01 - 0.1$, $\mu$ and $B\mu$.
In addition a $D$-term contribution is needed in order to obtain a phenomenologically acceptable theory~\cite{Raby:1998xr}.
The boundary conditions for the gauginos and scalars at the scale $M$ read
\begin{subequations}\label{eq:GMSBboundary}
\begin{eqnarray}
M_1 &=& \frac{3}{5} \frac{\alpha_1(M)}{4\pi}\Lambda\,
\left(1+\frac{20}{3} b^2\right)\;,  \\
M_2 &=& \frac{\alpha_2(M)}{4\pi}\Lambda\, \left(1+4 b^2\right)\;,  \\
M_3 &=& \frac{\alpha_3(M)}{\pi}\Lambda\, b^2 \;, \\
\widetilde{m}^2 &=& 2\, \Lambda^2\, \bigg\{C_1\, \left(\frac{\alpha_1(M)}{4\pi}\right)^2\left(\frac{3}{5}
               +\frac{2}{5}a^2 +4 b^2\right)
               + C_2\,\left(\frac{\alpha_2(M)}{4\pi}\right)^2
               \left(1 + 4b^2\right)
\nonumber \\
               & &\hphantom{2\, \Lambda^2\, \bigg\{}{}
               +  C_3\,\left(\frac{\alpha_3(M)}{4\pi}\right)^2
               \left(a^2 + 4b^2\right)\bigg\}
\end{eqnarray}
\end{subequations}
with $C_1=\tfrac{3}{5} Y^2$, $C_2=\tfrac{3}{4}$ for weak doublets and zero otherwise and
$C_3=\tfrac{4}{3}$ for color triplets and zero otherwise.
The additional $D$-term contributions to the scalar masses are parameterized as
\begin{equation}
\delta_D\, \widetilde{m}^2 ~=~ d \,Q_a^X \,M_2^2
\end{equation}
with $d$ an arbitrary parameter of order one and $Q_a^X$ the U(1)$_X$ charge of the field with label $a$.
In the following we will assume that $Q_a^X$ is 1 for the matter fields and -2 for the Higgs fields~\cite{Raby:1998xr}.

\begin{figure}[h!]
\centerline{\subfigure[$a=0.01$, $\tan\beta=10$, $A_0=0$, $d=0$.]{\includegraphics[width=7.5cm]{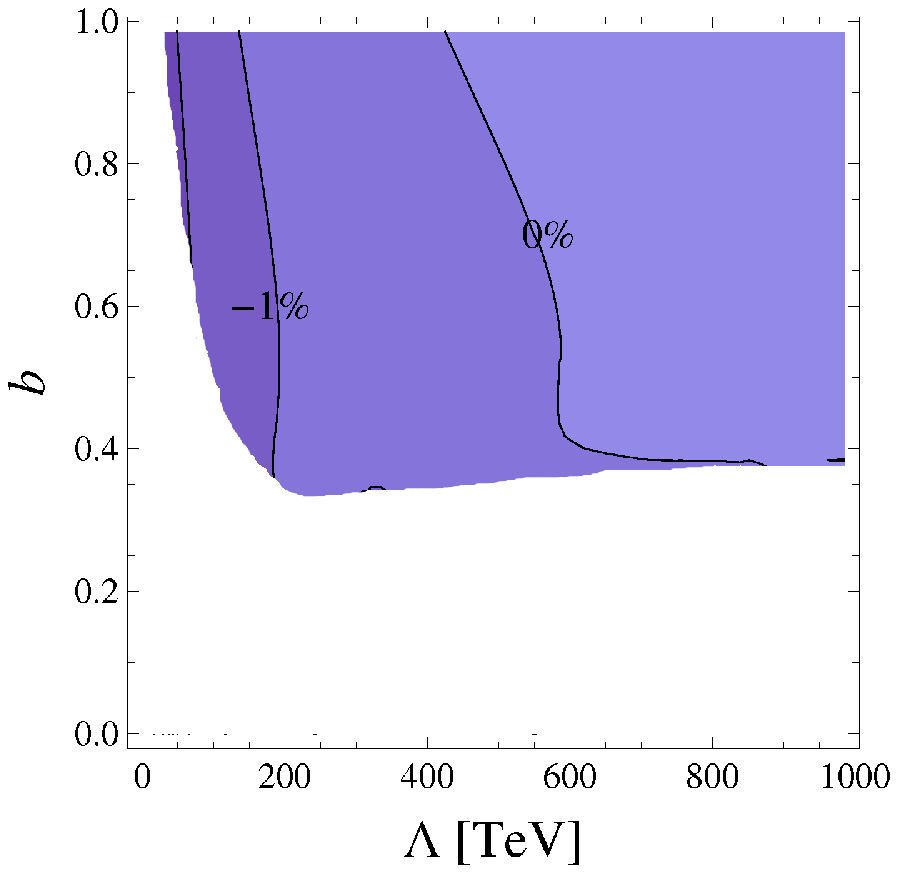}}
\subfigure[$a=0.01$, $\tan\beta=10$, $A_0=0$, $d=1$.]{\includegraphics[width=7.5cm]{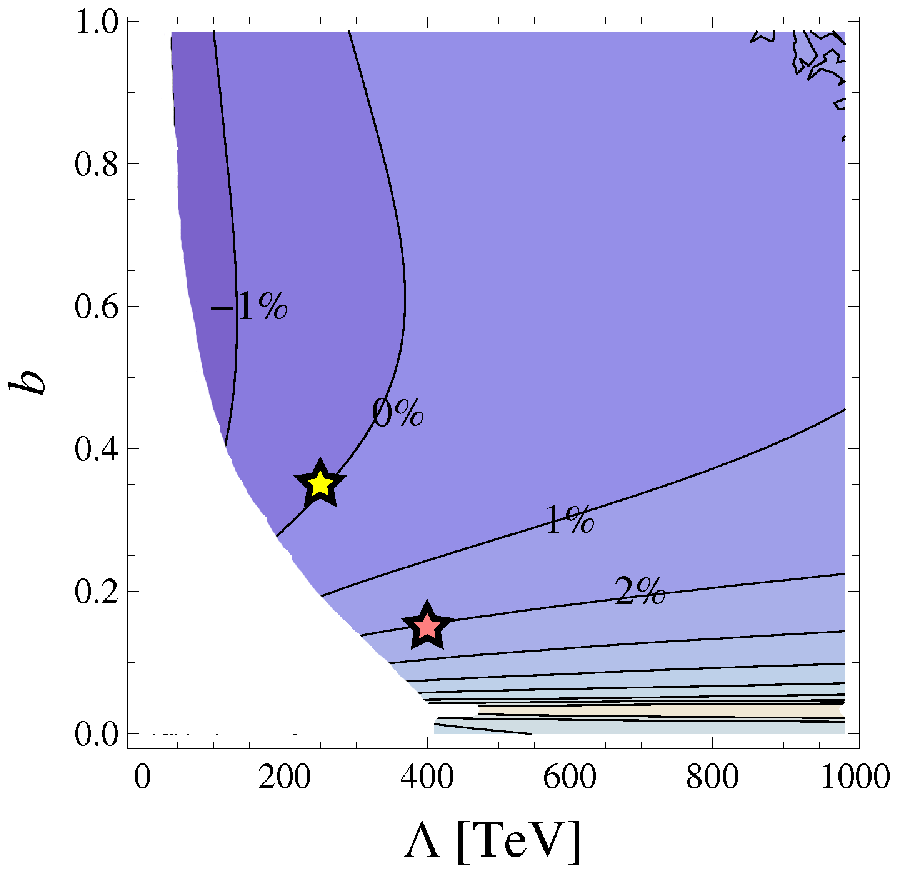}}}
\centerline{\subfigure[$a=0.01$, $\tan\beta=10$, $A_0=0$, $d=0$.]{\includegraphics[width=7.5cm]{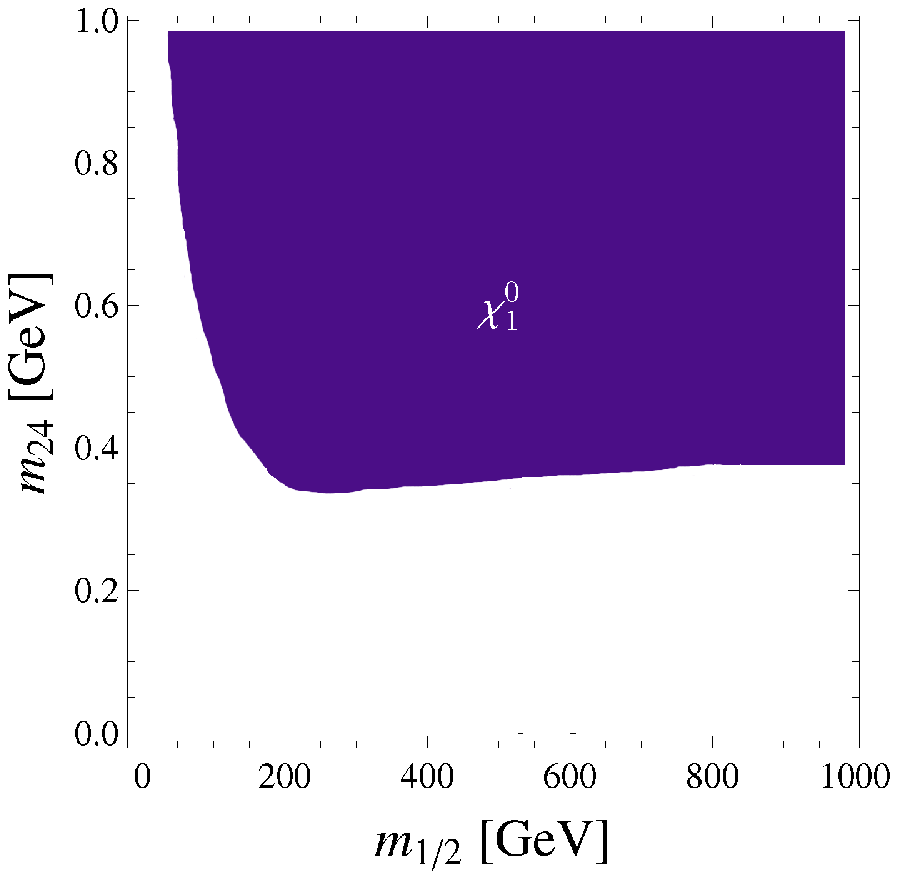}}
\subfigure[$a=0.01$, $\tan\beta=10$, $A_0=0$, $d=1$.]{\includegraphics[width=7.5cm]{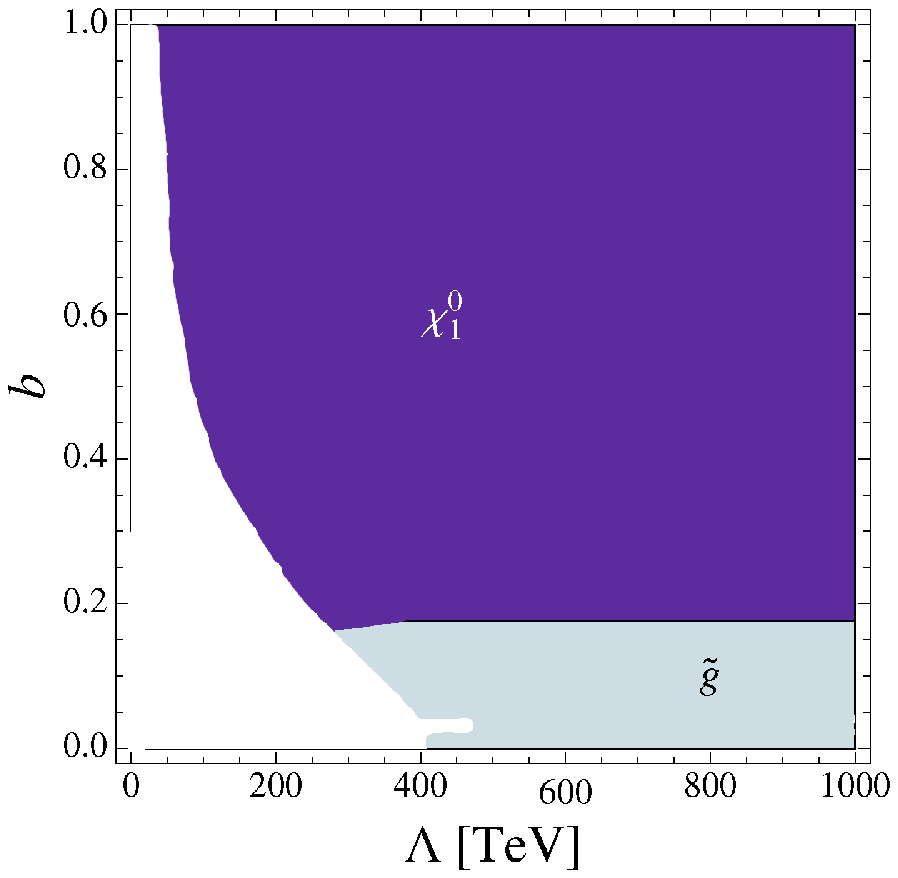}}}
\caption{Contours of different $\epsilon_3$ (upper panels) as well as the corresponding (N)LSPs (lower panels)
         for the GMSB with Higgs messenger mixing
         with $a=0.01$, $A_0=0$ and $\tan\beta=10$.
         The white region is excluded as before. The yellow and pink stars denote points in parameter space 
         for which we show the spectra in 
         table~\ref{tab:SampleSpectrum} and table~\ref{tab:SampleSpectrume2}.}
\label{fig:sraby}
\end{figure}

In figure~\ref{fig:sraby} we show again the contours of different
$\epsilon_3$ for the boundary conditions \eqref{eq:GMSBboundary} at
$M=M_\text{GUT}$ and $a=0.01$, $A_0=0$, $\tan\beta=10$ and $d=0$ or
$d=1$. In the region with small values of $b$ the gluino is the NLSP
with the gravitino being the LSP and  $\epsilon_3$ is positive.
A sample spectrum is shown in
table~\ref{tab:SampleSpectrume2}. Precision gauge coupling can be
achieved for larger values of $b$ where the neutralino is the NLSP.
A sample spectrum is shown in table~\ref{tab:SampleSpectrum}.

\section{Phenomenological implications}
\label{sec:Pheno}
\subsection{Prospects for the LHC}
Let us now briefly comment on possible phenomenological
implications of the requirement of precision gauge coupling unification. Sample
spectra for the considered models are shown in table~\ref{tab:SampleSpectrum}.
One common feature of the models which admit precision gauge coupling
unification seems to be that the gauginos are rather light while the scalars are
typically somewhat heavier.  It is therefore well possible that the gauginos are
within the kinematical reach of the LHC while the scalars are not.  At the LHC
one would then expect that supersymmetric particles are mainly produced via
gluino pair production,  resulting in a signature with at least four jets plus
missing energy. With an integrated luminosity of $100\,\mathrm{fb}^{-1}$ the LHC
could probe gluino masses up to $\sim 2\tev$~\cite{Wells:2003tf}.   As the
squarks are heavier than the gluinos, the gluinos can either three-body decay
via a virtual squark or two-body loop decay (for branching ratios see
e.g.~\cite{Barbieri:1987ed,Baer:1990sc}).  The latter process is usually
suppressed. As can be seen in table~\ref{tab:SampleSpectrum} the lighter stop is
the lightest among the squarks, implying that the gluino decays will produce a
high multiplicity of top quarks~\cite{Wells:2003tf}, if kinematically
accessible.  To summarize, squarks may be so heavy that they cannot be directly
produced, in which case their masses have to be reconstructed by investigating the
gluino decay processes, where they appear as intermediate states.

For positive $\epsilon_3$ the gluino is likely to be the lightest MSSM
superpartner,  in which case it may be long-lived.  Depending on the lifetime
the gluino hadronizes, travels macroscopic distances and might even escape the
detector  (see e.g.~\cite{Kilian:2004uj}). Potential signatures of such a
scenario could therefore be displaced vertices or metastable R-hadrons which
could be stopped in- or outside the detector. In any case the missing energy
signature for SUSY is not present in such a scenario.

\subsection{Further comments on scenarios with a low gluino mass}
As is well known, scenarios with low gluino masses can also be interesting in
the context of the little hierarchy problem. This is because the gluino mass has
the largest impact on the Higgs mass parameters (cf.~\cite{Kane:2002ap}).
Given a specific measure of ``fine-tuning'' one also finds that low gluino
masses increase naturalness (cf.~e.g.~\cite{Gogoladze:2009bd,Horton:2009ed} for
recent discussions).
On the other hand, one should also mention that a large gluino mass has
the virtue that the supersymmetric flavor problem gets ameliorated by
flavor-universal contributions from the gluino in the renormalization group
\cite{Dine:1990jd}.

\begin{table}[h]
\begin{center}
\begin{tabular}{|c|c|c|c|c|}
\hline
                                         & $m_0=m_{3/2}$          & $m_0=0.01\, m_{3/2}$    & non-singlet              & GMSB         \\
\hline
 $m_{h^0}$                               & 121~GeV                & 115~GeV                 & 115~GeV                  & 116~GeV\\
 $m_{H^0}$                               & 5.59~TeV               & 1.31~TeV                & 1.21~TeV                 & 1.39~TeV\\
 $m_{A^0}$                               & 5.59~TeV               & 1.31~TeV                & 1.21~TeV                 & 1.39~TeV\\
 $m_{H^+}$                               & 5.59~TeV               & 1.31~TeV                & 1.21~TeV                 & 1.39~TeV\\
 $m_{\widetilde{u}_{\mathrm{L}\,1/2}}$   & 7.37~TeV               & 1.55~TeV                & 1.57~TeV                 & 2.02~TeV\\
 $m_{\widetilde{u}_{\mathrm{R}\,1/2}}$   & 7.42~TeV               & 1.42~TeV                & 1.23~TeV                 & 1.62~TeV\\
 $m_{\widetilde{d}_{\mathrm{L}\,1/2}}$   & 7.41~TeV               & 1.55~TeV                & 1.57~TeV                 & 2.02~TeV\\
 $m_{\widetilde{d}_{\mathrm{R}\,1/2}}$   & 7.42~TeV               & 1.23~TeV                & 1.20~TeV                 & 1.53~TeV\\
 $m_{\widetilde{t}_{1}}$                 & 5.89~TeV               & 932~GeV                 & 743~GeV                  & 1.18~TeV\\
 $m_{\widetilde{t}_{2}}$                 & 6.66~TeV               & 1.36~TeV                & 1.41~TeV                 & 1.86~TeV\\
 $m_{\widetilde{b}_{1}}$                 & 6.66~TeV               & 1.22~TeV                & 1.19~TeV                 & 1.52~TeV\\
 $m_{\widetilde{b}_{2}}$                 & 7.39~TeV               & 1.15~TeV                & 1.40~TeV                 & 1.85~TeV\\
 $m_{\widetilde{e}_{\mathrm{L}\,1/2}}$   & 7.48~TeV               & 1.56~TeV                & 1.45~TeV                 & 1.81~TeV\\
 $m_{\widetilde{e}_{\mathrm{R}\,1/2}}$   & 7.49~TeV               & 1.42~TeV                & 1.08~TeV                 & 1.44~TeV\\
 $m_{\widetilde{\nu}_{\mathrm{L}\,1/2}}$ & 7.48~TeV               & 1.55~TeV                & 1.45~TeV                 & 1.81~TeV\\
 $m_{\widetilde{\tau}_{1}}$              & 7.45~TeV               & 1.55~TeV                & 1.07~TeV                 & 1.43~TeV\\
 $m_{\widetilde{\tau}_{2}}$              & 7.47~TeV               & 1.58~TeV                & 1.45~TeV                 & 1.81~TeV\\
 $m_{\widetilde{\nu}_{\tau_L}}$          & 7.46~TeV               & 1.55~TeV                & 1.44~TeV                 & 1.81~TeV\\
 $m_{\widetilde{g}}$                     & 128~GeV                & 793~GeV                 & 862~GeV                  & 828~GeV\\
 $m_{\chi_1^0}$                          & 117~GeV                & 583~GeV                 & 467~GeV                & 346~GeV\\
 $m_{\chi_2^0}$                          & 118~GeV                & 592~GeV                 & 644~GeV                & 922~GeV\\
 $m_{\chi_1^\pm}$                        & 118~GeV                & 591~GeV                 & 643~GeV                & 922~GeV\\
 $m_{\chi_3^0}$                          & 5.65~TeV               & 1.29~TeV                & 645~GeV                & 1.55~TeV\\
 $m_{\chi_4^0}$                          & 5.65~TeV               & 1.46~TeV                & 1.30~TeV                 & 1.55~TeV\\
 $m_{\chi_2^\pm}$                        & 5.68~TeV               & 1.29~TeV                & 1.30~TeV                 & 1.56~TeV\\
 $M_\mathrm{GUT}$                        & $2.59\cdot10^{16}$~GeV & $1.54\cdot10^{16}$~GeV  & $1.42\cdot10^{16}$~GeV   & $1.50\cdot10^{16}$~GeV\\
\hline
\end{tabular}
\caption{Sample spectra with vanishing $\epsilon_3$ for the different cases. For the case `$m_0=m_{3/2}$' (`$m_0=0.01 \, m_{3/2}$') we
         choose $m_{3/2}=7.5$~TeV (100 TeV), $\alpha=2.3$ (2.0) and $\tan\beta=10$. In the case `non-singlet' we have
         $m_{1/2}=848.5\gev$, $m_{24}=-500\gev$, $m_0=1 \tev$, $A_0=0$ and $\tan\beta=10$. For `GMSB' we have
         $a=0.01$, $b=0.32$, $A_0=0$, $\Lambda=250 \tev$, $d=1$ and $\tan\beta=10$.
         The lightest neutralino turns out to be mainly bino in the
         cases  `$m_0=m_{3/2}$', `non-singlet' and `GMSB' while it is mainly Higgsino in the case `$m_0=0.01 \, m_{3/2}$'.}
\label{tab:SampleSpectrum}
\end{center}
\end{table}

\begin{table}[h]
\begin{center}
\begin{tabular}{|c|c|c|}
\hline
                                         & $m_0=0.01\, m_{3/2}$      & GMSB   \\
\hline
 $m_{h^0}$                               & 117~GeV                   & 115~GeV\\
 $m_{H^0}$                               & 2.23~TeV                  & 1.60~TeV\\
 $m_{A^0}$                               & 2.23~TeV                  & 1.60~TeV\\
 $m_{H^+}$                               & 2.24~TeV                  & 1.61~TeV\\
 $m_{\widetilde{u}_{\mathrm{L}\,1/2}}$   & 2.63~TeV                  & 2.36~TeV\\
 $m_{\widetilde{u}_{\mathrm{R}\,1/2}}$   & 2.50~TeV                  & 1.67~TeV\\
 $m_{\widetilde{d}_{\mathrm{L}\,1/2}}$   & 2.64~TeV                  & 2.37~TeV\\
 $m_{\widetilde{d}_{\mathrm{R}\,1/2}}$   & 2.12~TeV                  & 1.52~TeV\\
 $m_{\widetilde{t}_{1}}$                 & 1.70~TeV                  & 1.07~TeV\\
 $m_{\widetilde{t}_{2}}$                 & 2.29~TeV                  & 2.18~TeV\\
 $m_{\widetilde{b}_{1}}$                 & 2.10~TeV                  & 1.51~TeV\\
 $m_{\widetilde{b}_{2}}$                 & 2.29~TeV                  & 2.17~TeV\\
 $m_{\widetilde{e}_{\mathrm{L}\,1/2}}$   & 2.86~TeV                  & 2.37~TeV\\
 $m_{\widetilde{e}_{\mathrm{R}\,1/2}}$   & 3.04~TeV                  & 1.81~TeV\\
 $m_{\widetilde{\nu}_{\mathrm{L}\,1/2}}$ & 2.86~TeV                  & 2.36~TeV\\
 $m_{\widetilde{\tau}_{1}}$              & 2.85~TeV                  & 1.80~TeV\\
 $m_{\widetilde{\tau}_{2}}$              & 3.02~TeV                  & 2.36~TeV\\
 $m_{\widetilde{\nu}_{\tau_L}}$          & 2.85~TeV                  & 2.36~TeV\\
 $m_{\widetilde{g}}$                     & 385~GeV                   & 301~GeV\\
 $m_{\chi_1^0}$                          & 963~GeV                   & 379~GeV\\
 $m_{\chi_2^0}$                          & 968~GeV                   & 1.14~TeV\\
 $m_{\chi_1^\pm}$                        & 974~GeV                   & 1.14~TeV\\
 $m_{\chi_3^0}$                          & 2.21~TeV                  & 1.63~TeV\\
 $m_{\chi_4^0}$                          & 2.76~TeV                  & 1.64~TeV\\
 $m_{\chi_2^\pm}$                        & 2.21~TeV                  & 1.64~TeV\\
 $M_\mathrm{GUT}$                        & $1.33 \cdot 10^{16} \gev$ & $1.39 \cdot 10^{16} \gev$\\
\hline
\end{tabular}
\caption{Sample spectra with $\epsilon_3=0.02$. In the case `$m_0=0.01 \,
		 m_{3/2}$' we choose $m_{3/2}=200\tev$ , $\alpha=1.615$ and
		 $\tan\beta=10$ while in the case `GMSB' we have $a=0.01$, $b=0.145$,
		 $A_0=0$, $\Lambda=400 \tev$, $d=1$ and $\tan\beta=10$. In both examples
		 the gluino is the lightest MSSM superparticle, such that dark matter
		 has to be composed of some other, probably very weakly coupled
		 particle(s).}
\label{tab:SampleSpectrume2}
\end{center}
\end{table}

\section{Summary}
\label{sec:summary}

We have shown that it is possible to have precision gauge coupling unification
in the MSSM without invoking high-scale thresholds. This can be simply achieved
in schemes in which the gaugino masses are non-universal at (or slightly below)
the GUT scale. Such non-universal gaugino masses are consistent with unified
theories. They occur, for instance, in settings in which the tree-level value of
the gaugino masses are suppressed such that quantum corrections, such as
`anomaly mediated' contributions, become important. They can also be obtained in
settings in which supersymmetry is broken by a field furnishing a non-trivial
GUT representation as well as in GMSB with Higgs-messenger
mixing. We also note that $\epsilon_3$ can be positive.

Interestingly, there is a correlation between low fine-tuning and precision
gauge coupling unification. Both features can be achieved by having a rather
light gluino. The emerging schemes have the virtue that it will be relatively
easy for the LHC to produce gluinos copiously. On the other hand, squarks may
turn out to be too heavy to be directly produced such that one would have to
reconstruct their masses by analyzing gluino decays. There is also the
possibility of displaced vertices from gluino decay as well as meta-stable
gluino bound states stopped in the detector.

\subsubsection*{Acknowledgments.}

We would like to thank  G.~Ross for useful discussions. S.R.\ would like to
acknowledge partial support from DOE grant DOE/ER/01545-883. This research was
supported by the DFG cluster of excellence Origin and Structure of the Universe
and the \mbox{SFB-Transregio} 27 "Neutrinos and Beyond" by Deutsche
Forschungsgemeinschaft (DFG). One of us (M.R.) would like to thank the Aspen
Center for Physics, where some of this work has been carried out, for
hospitality and support. K.S.-H.\ would like to thank LPSC Grenoble for
hospitality and support.

\bibliographystyle{ArXiv}
\bibliography{Orbifold}

\end{document}